\documentclass[preprint, 12 pt]{elsarticle}

\usepackage{graphicx}
\usepackage{hyperref}
\usepackage[utf8]{inputenc}
\usepackage{multirow}
\usepackage{array,multirow}
\usepackage{hyperref}
\usepackage{caption}
\usepackage{subcaption}
\usepackage[legalpaper, margin=1in]{geometry}

\usepackage{amssymb}
\usepackage{amsmath}

\journal{arXiv}
\begin{document}

\begin{frontmatter}


\title{C-Net: A Reliable Convolutional Neural Network for Biomedical Image Classification}



 \author[address1]{Hosein Barzekar\corref{corresponding author}}
\cortext[corresponding author]{Corresponding author}
\ead{barzekar@uwm.edu}

\author[address1,address2]{Zeyun Yu}
\ead{yuz@uwm.edu}

\address[address1]{Big Data Analytics and Visualization Laboratory, Department of Computer Science, University of Wisconsin-Milwaukee, Milwaukee, WI 53211, USA}
\address[address2]{Department of Biomedical Engineering, University of Wisconsin-Milwaukee, Milwaukee, WI 53211, USA}

\begin{abstract}
Cancers are the leading cause of death in many countries. Early diagnosis plays a crucial role in having proper treatment for this debilitating disease. The automated classification of the type of cancer is a challenging task since pathologists must examine a huge number of histopathological images to detect infinitesimal abnormalities. In this study, we propose a novel convolutional neural network (CNN) architecture composed of a Concatenation of multiple Networks, called C-Net, to classify biomedical images. The model incorporates multiple CNNs including Outer, Middle, and Inner. The first two parts of the architecture contain six networks that serve as feature extractors to feed into the Inner network to classify the images in terms of malignancy and benignancy. The C-Net is applied for histopathological image classification on two public datasets, including BreakHis and Osteosarcoma. To evaluate the performance, the model is tested using several evaluation metrics for its reliability. The C-Net model outperforms all other models on the individual metrics for both datasets and achieves zero misclassification. This approach has the potential to be extended to additional classification tasks, as experimental results demonstrate utilizing extensive evaluation metrics.
\end{abstract}

\begin{keyword}
Biomedical Image Classification \sep Deep Learning \sep Convolutional Neural Network\sep Histopathology \sep Computer-aided Diagnosis


\end{keyword}

\end{frontmatter}

\section{Introduction}
Cancers are the first or second leading causes of premature death in 134 out of 183 nations, and it ranks the third in 45 other countries. 4.5 million (29.8\%) of the 15.2 million world-wide premature deaths in 2016 were caused by cancers. Lung cancer is the most prevalent type with nearly 2.1 million new cases and 1.8 million deaths in 2018. Breast cancer is the most commonly diagnosed cancer in women with  2,088,849 new cases in 2018 and causing the world's largest cancer mortality in women (626,679 deaths in 2018) \cite{cancerstat, wild2020world}. 

An abnormal new tissue growth, known as a neoplasm, may be benign or malignant referring to the overall biologic behavior of a tumor rather than to its morphologic characteristics. The word "cancer" is commonly related to malignant tumors. In some conditions, however, benign tumors can be lethal \cite{rubin2014essentials}. Contingent upon the growth location, a cancer can have different clinical features including: rapid proliferation, diminished growth control, metastasis, and loss of contact inhibition in vitro \cite{rodwell2015harpers}. 

Numerous conventional techniques such as histological biopsy tests, CT-imaging, magnetic resonance imaging (MRI), bronchoscopy, magnetic resonance mammography (MRM) are some of the essential tools for diagnosing different types of cancer \cite{fass2008imaging, prabhakar2018current, geller2010osteosarcoma}. While biopsy-based technology can efficiently diagnose malignancy, major challenges remain in expediting clinical diagnosis from pathology imaging and in automated image processing \cite{wang2019pathology}. Automatically classifying histopathological images in computer-assisted pathology research is an imperative task. Because of the instability in appearance caused by the heterogeneity of diseases, tissue preparation, and staining process, deriving descriptive and concise information from histopathological images is considerably challenging \cite{feng2018deep}. The workload and complexity of histopathologic cancer diagnosis due to the emergence of precision medicine have increased significantly for pathologists, and a huge number of slides need to be examined for an incredibly long time before making any decision \cite{litjens2016deep}. 

Whole slide imaging(WSI) or virtual microscopy requires examining or digitizing glass slides to provide digital images necessary for automated image analysis being used in surgical pathology in regular practice \cite{pantanowitz2011review}. WSI has brought an opportunity to mitigate histopathology workload and assist pathologists to examine and quantify a large number of image slides. WSI combined with artificial neural networks and deep learning gives a tremendous confidence to improve the efficiency of diagnosing cancers \cite{litjens2016deep,lisboa2006use}. The accuracy of the models is growing continuously as the digital pathology is expanding and more datasets are becoming publicly available \cite{ibrahim2020artificial}.\\ 
In particular fields such as medical imaging, deep learning algorithms have outperformed human experts substantially \cite{schmidhuber2015deep}. For the prediction of complicated tasks, deep learning algorithms use massive datasets. In computer vision tasks where interpreting the images, deep learning has achieved a significantly high level of accuracy \cite{duggento2020deep}. CNN is a particular kind of neural network which are the most efficient for analyzing data in a grid-form topology. Convolution is a special linear operator for spatial and grid-like data such as digitized images. Convolutional networks are a basic neural network utilizing convolution rather than the matrix multiplication in at least one of its layers \cite{goodfellow2016deep}. In recent years, the deep learning architectures, particularly Deep Convolutional Neural Networks (DCNNs), are the most effective practices of analysis of medical imaging tasks such as diagnosis \cite{Talo_2019}.\\
DCNNs have gained a lot of attention and have been utilized for many classification tasks. The introduction of AlexNet \cite{krizhevsky2012imagenet} was a recent breakthrough in the field of machine and deep learning. The idea of having a deeper network has been the focus of many successful implementations of DCNNs including the idea of Networks inside Networks (NIN) \cite{lin2013network}. An example of this idea is the Inception \cite{szegedy2015going} network which achieved an outstanding performance in object detection and classification. Consequently, plenty of studies have adopted these implementations upon medical image classification, especially on breast cancer. However, in this work, we changed the focus on going deeper with CNNs by Concatenating multiple CNNs composed of Outer, Middle, and Inner networks. With this new architecture, which we call C-Net, several NINs are operating simultaneously on multiple networks.\\
The main contributions in this work can be summarized in the following aspects:
\begin{itemize}
    \item We propose a novel deep convolutional neural network, called C-Net, to surmount shortcomings in current classification schemes.
    \item We provide a comparative and successful application of the new architecture in the domain of cancer classification.
    \item  The model has been evaluated on two different datasets in terms of size, disease type, and image resolution: (1) BreakHis, a breast cancer dataset containing 7909 histopathology images with different magnification factors including 40X, 100X, 200X, and 400X, and (2) Osteosarcoma dataset composed of 1144 images, with the size of 1024$\times$1024 and 10X magnification.
\end{itemize}

The rest of the paper is organized as follows: Section. \ref{sec: RelatedWork} discusses related work on cancer classification. Section. \ref{Methology} provides a detailed explanation of the C-Net model which includes the datasets, pre-processing and the model architecture. The experimental results and the performance analysis are presented in Section. \ref{experimentaResult}. Finally, we present the conclusion and suggest future directions in Section. \ref{conclusion}. 

\section{Related Works} \label{sec: RelatedWork}
The advancements of CNNs are so substantial that many CNN-based models have surpassed human capacities in numerous fields \cite{litjens2017survey}. In classifying skin cancer, for example, DCNN has reached the dermatologist-level\ in identifying keratinocyte carcinomas versus benign seborrheic keratoses, and malignant melanomas versus benign nevi \cite{esteva2017dermatologist}. Successfully classifying and detecting lung cancer with high confidence at the radiologist level is another successful case of implementing DCNN which provides physicians an accurate diagnostic tool to detect and classify  pulmonary nodules into malignant or benign \cite{zhang2019toward}.

Several CNN architectures have been implemented on WSI. Cruz-Roa et al. employed a CNN model for invasive tumor detection on WSI. The classifiers were trained on 400 samples and then tested separately from The Cancer Genome Atlas on 200 cases. In contrast to manually annotated regions of invasive ductal carcinoma, their approach resulted in a dice-coefficient of 75.86\%, and negative predictive value of 96.77\% \cite{cruz2017accurate}. With Google's inception v3 model Chang et al did a pilot study on breast cancer for classifying histopathological images and achieved a value of 93\% in Area Under the Curve(AUC) \cite{chang}.

In another work in 2017, Wahab et al. \cite{wahab2017two} developed a CNN model to classify mitotic and non-mitotic nuclei in breast cancer on histopathology, which achieved a F-measure of 79\%. Similarly in \cite{roy2019patch}, the authors proposed a patch-based classifier through two approaches including one patch in one decision (OPOD) and all patches in one decision (APOD), to automate the classification of histopathological images by using the breast histology image dataset.

Fondon et al. utilized a computer-aided tool for automatic classification of tissue malignancy. The Support Vector Machine (SVM) was used as the kernel classifier, which achieved an accuracy of 75.8\% on their experiments \cite{fondon2018automatic}.

A structure deep learning model was proposed by Han et al \cite{han2017breast}, in which a distance constraint of feature space was proposed to articulate feature space similarities. The model achieved an average accuracy of 96\% on binary classifications on different magnification factors. Their model used the idea of transfer learning, in which the networks trained upon a large dataset and the weights were then transferred. Some other studies have been implemented using transfer learning for classifying histopathology images \cite{xu2017effect, talo2019automated,de2019double}. A different study was done by Celik et al. to identify the invasive ductal carcinoma using transfer learning. They used two pre-trained well-known models, ResNet-50 and DenseNet-161, for their experiments; and obtained F-scores of 92.38\% and 94.11\% for DenseNet-161 and ResNet-50 respectively \cite{celik2020automated}. 

Pratiher et al. \cite{pratiher2018grading} applied the Bidirectional LSTM(Bi-LSTM) on manifold encoded histopathological image of the quasi-isometric topological space as an automated categorization framework for breast cancer classification. Morphological dynamics and contextual feature space semantic constraints have been utilized in their method through Bi-LSTM on histopathological manifolds.

In another model called DMAE \cite{feng2018deep}, a patch-based deep learning method comprises of two stages: (1) an end-to-end deep neural network was trained with histopathology images, and (2) previously unseen patches of the test images were used for breast histopathology image classification with the features learned from the proposed model.

Correct classification and detection of breast cancer in terms of malignancy and benignancy have been the focus of many studies \cite{araujo2017classification,spanhol2017deep, deniz2018transfer, awan2018context, wahab2019transfer}. Another study that was done in this regard was by Khan et al. \cite{khan2019novel}, in which they proposed a framework for extracting low-level features on cytology images using transfer learning. The results of their study have been observed to outclass other deep learning models for detecting and classifying breast cancer in terms of accuracy with an average of 97.67\% in cytology images. Guata et al. \cite{gupta2019partially} implemented partially-independent framework to explore the features on a multi-layer ResNet model for classification of histopathology images. The average accuracy of the four categories of BreakHis dataset was 94.66\%. 

An integration of multi-layer features from a ResNet model called Partially-Independent Framework was proposed by Gupta and  Bhavsar \cite{gupta2019partially} for breast cancer histopathological image classification. Since the discriminative features are not learned by all the layers, they decided to choose the optimal subset of layers based upon an information theoretic measure (ITS). They used XGBoost method for dimensionality reduction, and SVM with two polynomial kernels as their classifier framework. The highest accuracy was obtained by using 40X images at the rate of 97\%.


In an attempt to classify different histopathological images of Osteosarcoma into the necrotic tumor, viable tumor, and non-tumor, Arunachalam et al. employed a combination of machine and deep learning models for the task. Their model achieved an accuracy of 93.3\%, and 92.7\% for necrotic, 95.3\% for viable, and 91.9\% for non-tumor \cite{arunachalam2019viable}. The most recent study was done by Fu and Xue \cite{fu2020deep} for assessing viable and necrotic tumors on Osteosarcoma. The DS-Net they used was a combination of  an auxiliary supervision network (ASN) and a classification network. Features extracted by ASN were used for accurate classification. Their model achieved an overall accuracy of 94.5\%, and the class-based categories with accuracies of 92.2\%, 93.6\%, and 97.7\% for non-tumor, necrotic tumor, and viable tumor respectively. Anisuzzaman et al., performed four binary classifications between three osteosarcoma tumor types: non-tumor (NT), necrotic tumor (NCT), and viable tumor (VT) using transfer learning with VGG19 and InceptionV3 networks. They achieved the best results using VGG19 networks in all four binary classifications with the highest accuracy of 95.65\% on the NCT vs. NT category \cite{ANISUZZAMAN2021102931}.



\section{Methodology} \label{Methology}

\subsection{Datasets}
\subsubsection{BreakHis}
\begin{figure}
    \centering
    \includegraphics[width=12cm,height=7cm]{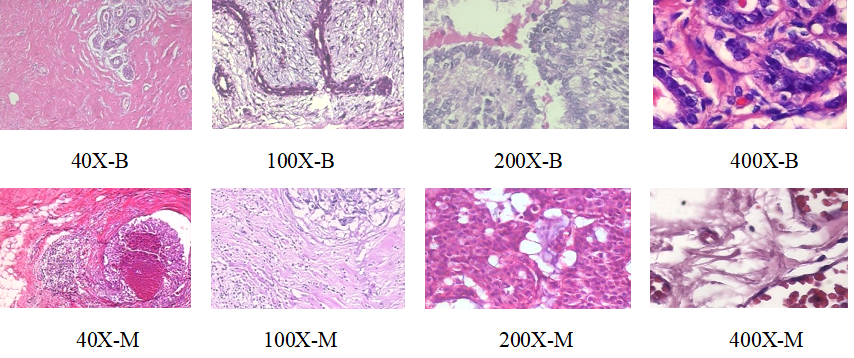}
    \caption{Sample Images from the BreakHis dataset; Top row shows several examples of benign tumors with different magnification factors, and the bottom row shows malignant tumors.}
    \label{fig:datasetBreakHis}
\end{figure}

The Breast Cancer Histopathological Image Classification (BreakHis) is a freely available public dataset (https://web.inf.ufpr.br/vri/databases/breast-cancer-histopathological-database-breakhis/),  composed of 7909 images, including 2480 benign and 5429 malignant cases,  with different magnification factors including 40X, 100X, 200X, and 400X. The size of the Images is 700$\times$460 with a three-channel RGB at each pixel \cite{spanhol2015dataset}. \autoref{tab:breakHis} shows the distribution of the BreakHis dataset by four magnification factors: 40X, 100X, 200X, and 400X of Benign and Malignant Cancer. \autoref{fig:datasetBreakHis} displays sample images of all types and magnifications.
\subsubsection{Osteosarcoma dataset}
\begin{figure}
    \centering
    \includegraphics[width=12cm,height=4cm]{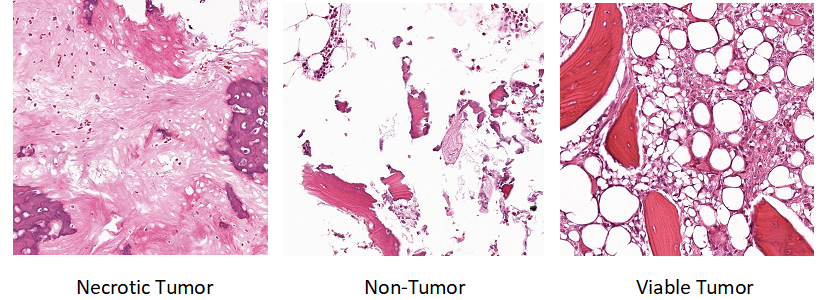}
    \caption{Sample Images from the OsteoSarcoma dataset including, Necrotic Tumor, Non-Tumor, Viable Tumor.}
    \label{fig:datasetOSteosarcoma}
\end{figure}

Another entirely different dataset to test our methods is composed of Hematoxylin and Eosin(H\&E) stained Osteosarcoma histology images, provided by the University of Texas Southwestern Medical Center. The samples were obtained from the pathology reports of the Osteosarcoma resection of 50 patients. The dataset contains 1144 images of 1024$\times$1024 pixels and 10X magnification, including 536, 263, and 345 images for non-tumor (NT), necrotic tumor (NCT), and viable tumor (VT) respectively \cite{osteodataset}. Among the 345 viable tumor images, 53 images are discarded from the experiments because they labeled with both viable and necrotic tumors. \autoref{fig:datasetOSteosarcoma} shows some sample images from the Osteosarcoma dataset.

\begin{table}[htbp]
  \centering
  \caption{The Distribution of the BreakHis dataset by Magnification Factors and Categories}
    \begin{tabular}{p{2cm}ccccc} \hline
    \multirow{2}[0]{*}{Class} & \multicolumn{4}{c}{Magnification} &  \multirow{2}[0]{*}{Total} \\ \cline{2-5}
          & 40X   & 100X  & 200X  & 400X  &  \\ \hline
    Benign & 625   & 644   & 623   & 588   & 2480 \\
    Malignant & 1370  & 1437  & 1390  & 1232  & 5429 \\ \hline
    Total  & 1995  & 2081  & 2013  & 1820  & 7909 \\ \hline
    \end{tabular}%
  \label{tab:breakHis}%
\end{table}%

\subsection{Pre-processing}
Machine/Deep learning models are inherently data-hungry, posing great challenges in the medical imaging field due to the lack of big data in training a new model. By generating artificial data and adding them into the training set, data augmentation provides a way to mitigate the obstacles with small datasets size \cite{shorten2019survey}.To this end, all the image intensities in our experiments are first scaled to the range between 0 to 1. The augmentation techniques that have been applied to them include flipping horizontally and vertically, shearing with a factor of 0.2, zooming by 0.2, height and width shifting with a factor of 0.2, and sequential rotation by 40 degrees.

\subsection{The C-Net architecture}
In developing our new architecture, called C-Net, different networks are integrated to fulfill several goals. (1) Using the same architecture in the Outer networks results in a more stable and reliable way of feature extracting, and using the same filter size in convolutions layers (3$\times$3), inspired by VGG-19 model \cite{simonyan2014very}, provides a better feature extraction compared to other implementations with different filter sizes. (2) The shortcomings of one network is compensated by another since the networks work in parallel and features are extracted by different networks at different times instead of going directly into the \textit{FC} layers. (3) The proposed model has fewer parameters to be trained. Compared to most successful architectures, the proposed model has fewer parameters, typically less than 30M with an image size of 224$\times$224 pixels.

\begin{figure}
    \centering
    \includegraphics[width=17cm,height=10cm]{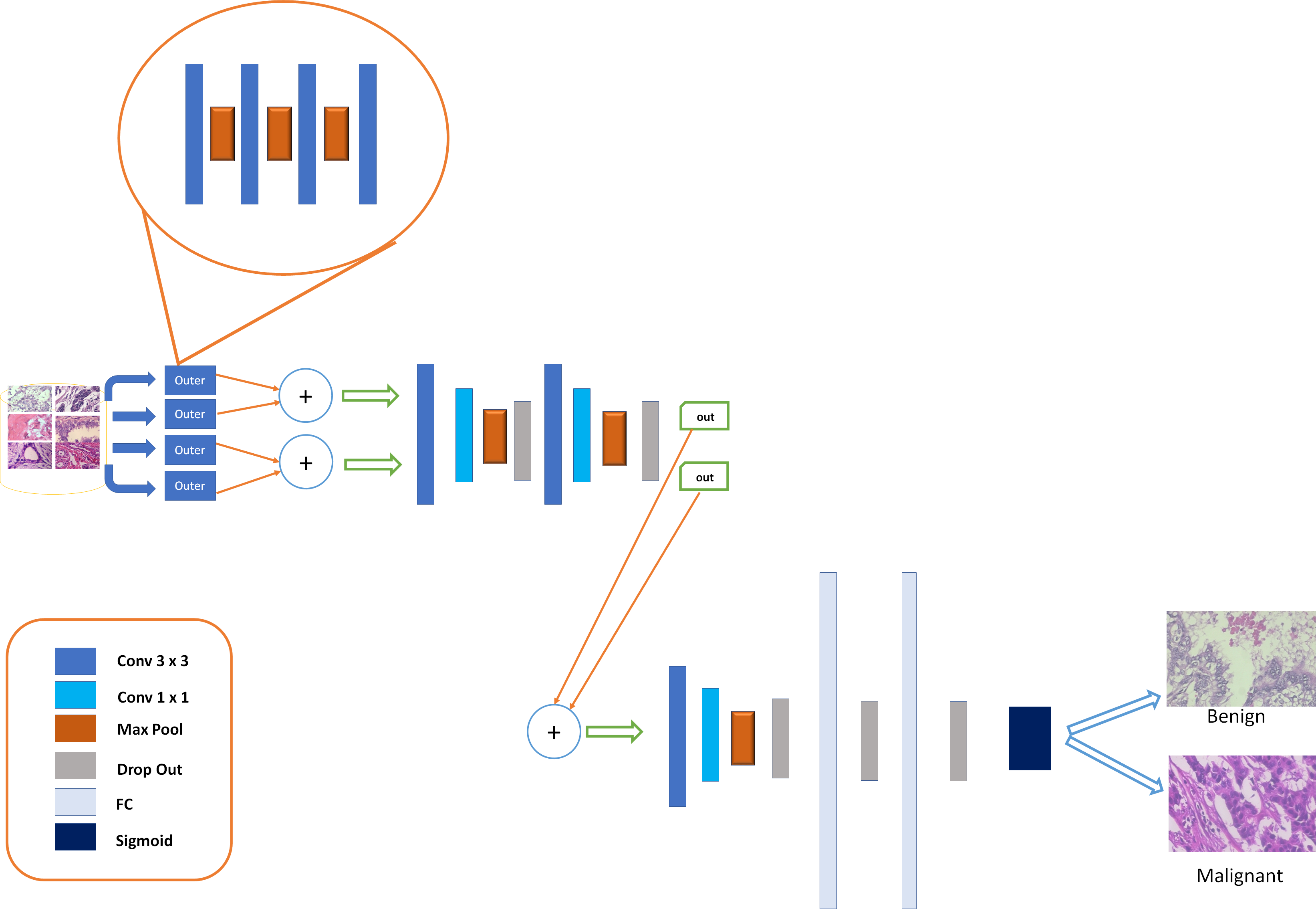}
    \caption{C-Net Architecture; Images feed into the Outer networks simultaneously, then the outputs get Concatenated to go through the Middle networks, and finally the Inner network takes the Concatenated outputs from the Middle networks.}
    \label{fig:Model}
\end{figure}

As shown in \autoref{fig:Model}, the C-Net comprises of three main parts, the Outer, Middle, and Inner networks. The Outer one is composed of four CNNs extracted from VGG19, and works as a feature extractor. The architecture of each of the outer networks is as follows: first, the images go into the input layer in all of the outer networks simultaneously followed by several convolutional layers, followed by a max-pooling layer in the first \textit{block}. Here, \textit{block} refers to a combination of several convolutional layers followed by a pooling layer.  The number of filters in the first block is 64 with a size of 3$\times$3  and the same padding, and the filter size of max-pooling is 2$\times$2 with a stride of 2. The same structure is being repeated with the same order for three additional \textit{blocks}, except that the number of filters gets multiplied by 2 but not the final \textit{block}. To prevent further reduction of the final output, the max-pooling layer has been dropped for the final \textit{block}. Rectified Linear Unite(\textit{ReLU}) is applied upon the convolution layers as the activation function.

The returning output features are then Concatenated, as shown by $\oplus$ sign in \autoref{fig:Model}, which is one of the important operations in the C-Net model shown in \autoref{eqConcatenation}. This operation is being applied two by two upon all of the Outer networks' output. 

\begin{equation}
\label{eqConcatenation}
\begin{split}
f(y, w) = \left((y_{m, n, c_i}) \oplus (w_{m, n, c_j})\right)\\
= X_{m, n, c_i+c_j}
\end{split}
\end{equation}
where \textit{(m,n)} are the dimensions of networks' outputs, y and w are the feature maps of different networks, $(c_i,c_j)$ are the  number of the channels on each output, $\oplus$ is the Concatenation with respect to the feature map axis and channel, and finally \textit{X} is the result of the Concatenation operation, which would be the input for Middle networks. 

Features extracted from Outer networks serve as the input for the Middle networks. The structure of the Middle networks is as follows: four convolutional layers on top of each other, each with a filter size of 3$\times$3, the same padding, and 256 filters. To overcome the model's complexity and reduce the feature maps, 1$\times$1 convolution has been placed upon the previous convolutions. NIN, 1$\times$1, is a fully connected network that is being applied upon the feature map, inspired by Lin et al. \cite{lin2013network}. Subsequently, a max-pooling layer with a stride of 2 has been placed after the 1$\times$1 convolutional layer to complete the first \textit{block} of the Middle networks. The same architecture is repeated in the second \textit{block} of the Middle networks. The activation function on all the convolutional layers is \textit{ReLU}. The outputs (feature maps) from the Middle networks are going to be Concatenated, (see \autoref{eqConcatenation}), and serve as the input for the Inner network. To generate efficient feature descriptors, we make sure that each network contains the max-pooling layer.

Drop out, which is randomly turning off some units of the layers, is a regularization technique and has an extreme effect of preventing the network from overfitting. It has been applied upon each \textit{block} of the Middle networks.

Finally, the Inner network takes the features returned by the Middle networks as the input. The Inner network contains only one \textit{block} which has the following structure: two convolutional layers with a filter size of 3$\times$3 and stride of 1, the same padding, and 256 filters. Additionally, we have a 1$\times$1 convolutional layer with same configurations followed by a max-pooling layer with the size of 2$\times$2, and stride of 2. In the Inner network the \textit{ReLU} is also used as the activation function. 

The max-pooling layer returned from the Inner network is flattened out into a vector connected to a (\textit{FC}) layer containing 1024 units and connected to another \textit{FC} layer with an equal number of units. Drop out has been applied upon both of the \textit{FC} layers. Finally, sigmoid is used as the activation function, as shown in \autoref{eqSigmoid}, at the output layer with two nodes, benign and malignant, as shown in \autoref{fig:Model}.

\begin{equation}
\label{eqSigmoid}
\begin{split}
z = w^Tx + b\\
\hat{y} = Sig(z)
 =\frac{e^z}{e^z+1}
\end{split}
\end{equation}
\textit{z} here is the dot product of filter \textit{w} with a chunk of the image with same size of the filter, and \textit{b} is the bias.

The loss function for the proposed model is a cross-entropy function represented as follow: 

\begin{equation}
\label{eqlossFunction}
\begin{split}
L(\hat{y}, y)=-\left(\sum_{i=1}^{N}y_{i}\log \hat{y}_{i}+(1-y_{i})\log(1-\hat{y}_{i})\right)
\end{split}
\end{equation}
where $y_i$ is the $i^{th}$ label y of N classes, and \^{y} is the $i^{th}$ element of the model's output \^{y}.

\section{Experimental Results and Analysis} \label{experimentaResult}
\subsection{Setup}
This section presents the results on two different histopathological datasets; both have been split into training, validation, and test images with a ratio of 70\%, 15\%, and 15\% respectively. For the BreakHis dataset, two-class classifications have been performed on different magnification  levels, including 40X, 100X, 200X, and 400X. The configuration for the Osteosarcoma dataset on two-class settings are as follows: a) NCT versus NT, b) NCT versus VT, and c) NT versus VT. 

The parameters setup for running the experiments are as follow: 
\begin{itemize}
    \item [--]Images are resized to 224$\times$224 and 375$\times$375 for BreakHis and Osteosarcoma respectively.
    \item[--]Adam algorithm is used for optimization. 
    \item[--]Different batch sizes including, 16, 32, and 64 are used.
    \item[--]Binary cross entropy is chosen as the loss function. 
\end{itemize}
The model has been implemented in Python, employing two well-known deep learning platforms, TensorFlow \cite{abadi2016tensorflow} and Keras \cite{chollet2015keras}, as its architecture. The experimental environment is run on Ubuntu 18.04, trained and tested on an Nvidia GeForce RTX 2080Ti GPU platform.

In order to measure the performance of the model several metrics are used. Precision (or Positive Predictive Value (PPV)) shows all the instances predicted as cancer (y=1), what fraction of them are truly cancerous. Negative Predictive Value (NPV), \autoref{eqNPV}, shows the probability of all the instances that the model predicted as negative. Recall (or Sensitivity), \autoref{eqRe}, tells what fraction of all the actual cancer instances the model detected as cancer. Specificity, \autoref{eqSpecificity}, is the capacity of the model to correctly find no cancer among non-cancerous. Matthews correlation coefficient (MCC), \autoref{eqMCC}, is another evaluation metric which produces high score when the model has achieved a superior performance on all the categories of confusion matrix including True Negative (TN), True Positive (TP), False Negative (FN), and False Positive (FP), as more unbiased and reliable than  accuracy (\autoref{eqAcc}) and F1-score (\autoref{eqF1-score}), in the presence of imbalance datasets \cite{chicco2020advantages}. The following formulas give the definitions of the aforementioned metrics from the confusion matrix.

\begin{equation}
\label{eqAcc}
\begin{split}
Accuracy =\frac{TP+TN}{TP+TN+FN+FP}
\end{split}
\end{equation}

\begin{equation}
\label{eqPre}
\begin{split}
Precision(PPV) =\frac{TP}{TP+FP}
\end{split}
\end{equation}

\begin{equation}
\label{eqNPV}
\begin{split}
NPV =\frac{TN}{TN+FN}
\end{split}
\end{equation}

\begin{equation}
\label{eqRe}
\begin{split}
Recall(Sensitivity) =\frac{TP}{TP+FN}
\end{split}
\end{equation}

\begin{equation}
\label{eqSpecificity}
\begin{split}
Specificity =\frac{TN}{TN+FP}
\end{split}
\end{equation}

\begin{equation}
\label{eqF1-score}
\begin{split}
F1-score =2\times\frac{\textit{Precision}\times\textit{Recall}}{\textit{Precision}+\textit{Recall}}
\end{split}
\end{equation}

\begin{equation}
\label{eqMCC}
\begin{split}
MCC =\frac{({TP}\times{TN})-({FP}\times{FN})}{\sqrt{(TP+FN)\times(TP+FP)\times(FP+TN)\times(TN+FN)}}
\end{split}
\end{equation}

\begin{figure}
    \centering
    \includegraphics[width=12cm,height=10cm]{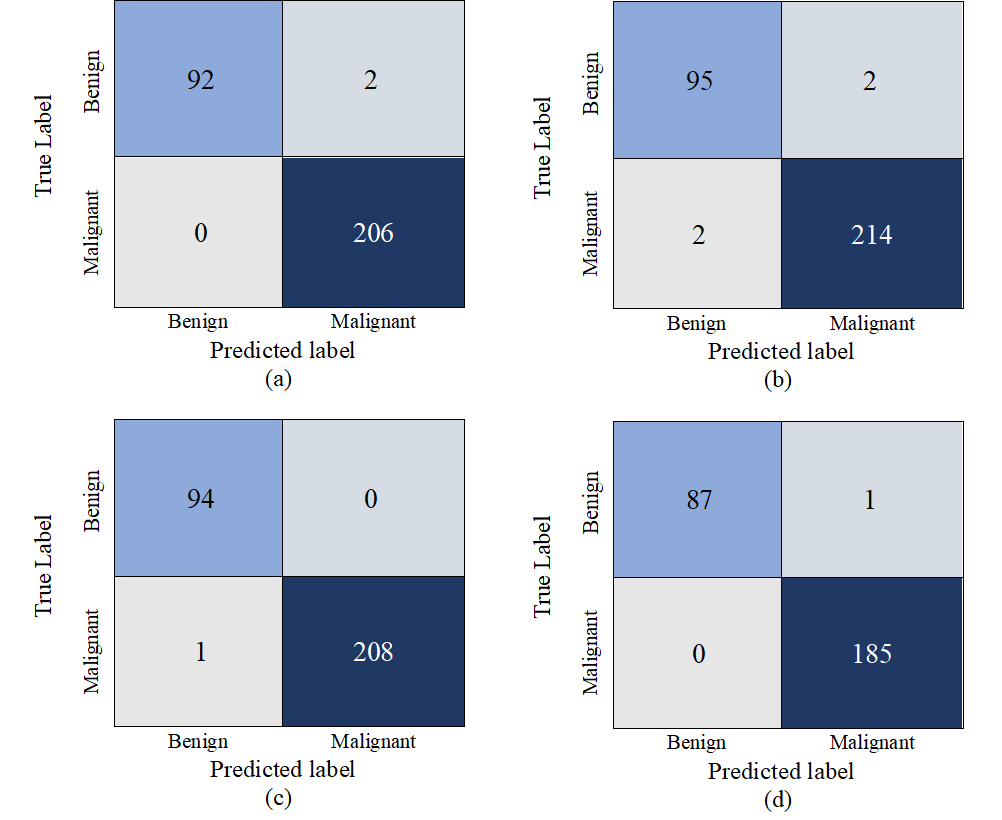}
    \caption{The confusion matrix of the C-Net on different magnification factors, \textbf{(a)} 40X images, \textbf{(b)} 100X images, \textbf{(c)} 200X images, \textbf{(d)} 400X images, of the test data on BreakHis dataset.
    }
    \label{fig:confusionMat_BreakHis}
\end{figure}

\begin{figure}
    \centering
    \includegraphics[width=15cm,height=5cm]{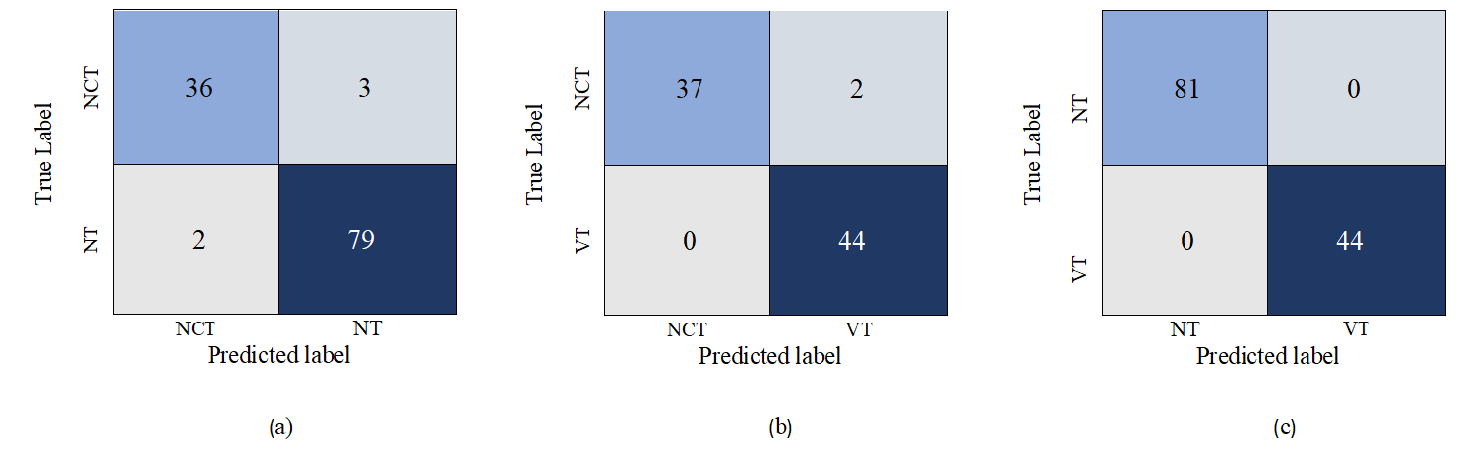}
    \caption{The confusion matrix of the C-Net on different classification categories; (a) NCT vs NT, (b) NCT vs VT, (c) VT vs NT}
    \label{fig:confusionMat_Osteosarcoma}
\end{figure}


\begin{table}[htbp]
  \centering
  \caption{Performance of the C-Net using different evaluation metrics in different magnification factors on BreakHis dataset; The values are in (\%).}
    \begin{tabular}{cccccc}
    \hline
    \multirow{2}[0]{*}{Evaluation Metrics} & \multicolumn{4}{c}{Magnificaion Factors} & \multirow{2}[0]{*}{Average} \\ \cline{2-5}
          & 40X   & 100X  & 200X  & 400X  &  \\ \hline
    Accuracy & 99.33 & 98.72 & 99.67 & 99.63 & 99.34 \\
    Precision(PPV) & 99.03 & 99.07 & \textbf{100}   & 99.62 & 99.43 \\
    NPV   & \textbf{100}   & 97.94 & 98.95 & \textbf{100}   & 99.22 \\
    Recall(\textit{Sensitivity}) & \textbf{100}   & 99.07 & 99.52 & \textbf{100}   & 99.65 \\
    Specificity & 97.87 & 97.94 & \textbf{100}   & 98.86 & 98.67 \\
    F1-score & 99.51 & 99.07 & 99.76 & 99.73 & 99.52 \\
    MCC   & 98.45 & 97.01 & 99.23 & 99.16 & 98.46 \\ \hline
    \end{tabular}%
  \label{tab:Metrics breakHis}%
\end{table}%

\subsection{Results}
The experimental results are composed of three parts. First, we present the results of the C-Net model on the BreakHis dataset upon different magnification factors (40X, 100X, 200X, 400X) using various evaluation metrics. Secondly, we show the classification results obtained by C-Net on the Osteosarcoma dataset. Finally, we demonstrate a comparative analysis of our model with previous models carried out upon the same dataset. Tables and matrices demonstrate the performance of the model on the test set. 

Different values in \autoref{tab:Metrics breakHis} are derived from confusion matrices presented in \autoref{fig:confusionMat_BreakHis}. \autoref{tab:osteo results} represents the results on three different settings of binary classification including, NCT vs NT, NCT vs VT, and ultimately VT vs NT. Furthermore, the confusion matrices for this dataset are given in \autoref{fig:confusionMat_Osteosarcoma}. A comparison between different architectures and the C-Net model on breast cancer classification using the BreakHis dataset is shown in \autoref{tab:comparisonBreakHis}.


\begin{table}[htbp]
  \centering
  \caption{Performance of the C-Net using different evaluation metrics in different classification settings Osteosarcoma dataset; NT(Non-Tumor), NCT(Necrotic Tumor), VT(Viable Tumor). The values are in (\%).}
    \begin{tabular}{crccc}
    \hline
    \multirow{2}[0]{*}{Evaluation Metrics} & \multicolumn{3}{c}{Different Classification settings} & \multirow{2}[0]{*}{Average} \\ \cline{2-4}
    
          & \multicolumn{1}{c}{NCT vs NT} & NCT vs VT & \multicolumn{1}{l}{VT vs NT} &  \\ \hline
    Accuracy & 95.83 & 97.59 & \textbf{100}   & 97.81 \\
    Precision(\textit{PPV}) & 94.74 & \textbf{100}   & \textbf{100}   & 98.25 \\
    \textit{NPV} & 96.34 & 95.65 & \textbf{100} & 97.33\\
    Recall(\textit{Sensitivity}) & 92.31 & 94.87 & \textbf{100}   & 95.73 \\
    Specificity & 97.53 & \textbf{100} & \textbf{100} & 99.18\\
    F1-score & 93.51 & 97.37 & \textbf{100}   & 96.96 \\
    \textit{MCC}   & 90.46 & 95.26 & \textbf{100}   & 95.24 \\ \hline
    \end{tabular}%
  \label{tab:osteo results}%
\end{table}%

\begin{table}[htbp]
  \centering
  \caption{Comparison of the success of different architectures with C-Net model tested on BreakHis dataset; The values are in (\%).}
    \begin{tabular}{cp{1.5cm}ccccc} \hline
    \multirow{2}[0]{*}{Architectures} & \multirow{2}[0]{*}{year} & \multicolumn{4}{c}{Magnification Accuracies in perc} & \multirow{2}[0]{*}{Avg.Acc } \\ \cline{3-6}
          &       & 40x   & 100x  & 200x  & 400x  &  \\ \hline
    CSDCNN\cite{han2017breast} & 2017  & 95.9  & 96.9  & 96.7  & 94.9  & 96.1 \\
    IRRCNN\cite{alom2019breast} & 2018  & 97.95 & 97.57 & 97.32 & 97.36 & 97.55 \\
    DMAE+SM\cite{feng2018deep} & 2018  & 94.43 & 93.15 & 99.36 & 94.86 & 95.45 \\
    Bi-LSTM\cite{pratiher2018grading} & 2018  & 96.2  & 97.2  & 97.1  & 95.4  & 96.48 \\
    BreastNet\cite{tougaccar2020breastnet} & 2019  & 97.99 & 97.84 & 98.51 & 95.88 & 97.55 \\
    Partially independent\cite{gupta2019partially} & 2019  & 97    & 96.1  & 94.69 & 90.85 & 94.66 \\
    Sharma et al.\cite{sharma2020novel} & 2020  & 97.4  & 98.6  & 97.7  & 96.8  & 97.63 \\
    \textbf{C-Net} & 2020  & \textbf{99.33} & \textbf{98.72} & \textbf{99.67} & \textbf{99.63} & \textbf{99.34} \\
    \hline
    \end{tabular}%
  \label{tab:comparisonBreakHis}%
\end{table}%

\subsection{Discussion} \label{discussion}
 Accurately diagnosing histopathological images is an extremely time-consuming process and needs expertise to manually go through 1000 slides to detect abnormalities, and it is prone to error. Early diagnosis of cancer could have a significant impact on the outcome of the disease and could save lives, and hence it has been the focus of many studies in cancer-related research \cite{sun2020deep, azer2019challenges, zheng2020deep}. Computer-aided diagnosis (CAD) systems, especially DCNN, aim to help physicians to diagnose the presence of cancer better and faster, and CAD has a transcendent advantage over traditional machine learning techniques \cite{chougrad2018deep}. 

Independent of the population of interest subjected to the test, sensitivity and specificity are usually employed for evaluating clinical tests \cite{lalkhen2008clinical}.  As presented in \autoref{tab:Metrics breakHis}, the C-Net model achieves 100\% performance  for these two metrics on different magnification factors including 40X, 100X, and 400X on the BreakHis dataset, which outperforms all the previous studies using these metrics. 

Sensitivity is the probability that the model predicts positive given the patient has the disease or $P(+|Disease)$, and likewise specificity is the probability that the model predicts negative given the patient does not have the disease, $P(- | Normal)$. Despite the popularity of these evaluation metrics, their use is not realistic for a clinician in determining the probability of illness in individual patients \cite{akobeng2007understanding}. Clinicians may want to know the probability that the patient has the disease or not when the model predicts positive or negative. Hence, PPV which is the $P(Disease | +)$, and NPV is $P(Normal | -)$ are two other measurements. As shown in \autoref{tab:Metrics breakHis} the C-Net achieves 100\% PPV for 200X, and also 100\% NPV for two different magnification factors, 40X and 400X, which corroborate the model as a robust classifier.

The lowest precision achieved by the proposed model as shown in \autoref{tab:Metrics breakHis} is 99.03\% for 40X images, and the rate increases as the magnification increases. Furthermore, the C-Net model achieves the 100\% recall rate on two magnification factors, 40X and 400X. Precision and recall alone cannot explicitly show the performance of a classifier. However, F1-score in \autoref{eqF1-score}, is when these two metrics are combined as their harmonic mean. The higher the score is the closer precision and recall are to each other. The model attains $>99\%$ of F1-score on all the magnification factors, and the average score of 99.52\%,  which compared to the average F1-score in \cite{sharma2020novel} with a rate of 97.67\% is a significant improvement and shows its robust performance for classifying medical images.


MCC in \autoref{tab:Metrics breakHis} is the next metric for evaluating a binary classification model. Even though the accuracy and F1-score are the most popular metrics in binary classification tasks, these metrics could be overly optimistic and inflate the results \cite{chicco2020advantages}. The C-Net model achieves the highest score of 99.23\% for MCC using 200X images, and the average of all MCC scores in different magnification factors is 98.46\%, which exceeds even the average accuracy of all the models presented in \autoref{tab:comparisonBreakHis}. This high performance on MCC value shows that Concatenating multiple networks results in better feature extraction and fewer misclassifications, with only 8 misclassified cases in all magnification factors combined, as is shown in the confusion matrices in \autoref{fig:confusionMat_BreakHis}.

 The confusion matrices allow us to have a detailed comparison of the models' performances. The highest MCC score, 99.23\%, by the C-Net model is 2.63\% greater than the highest MCC value based on the confusion matrix in \cite{tougaccar2020breastnet}. Additionally, the highest accuracy and F1-score of their model was attained on 200X images with a rate of 98.51\% and 98.28\% respectively. In the C-Net model, however, the highest accuracy (99.67\%) and F1-score (99.76\%) are obtained on the same magnification. Another evaluation metric for our comparison is sensitivity or recall. The highest sensitivity rate, 98.70\%, was achieved in \cite{tougaccar2020breastnet} on the same magnification factors, compared to our model that achieved the rate of 100\% on two magnification levels, 40X and 400X.   


\autoref{tab:comparisonBreakHis} illustrates a detailed comparison of accuracy achieved by previously successful models with our model on BreakHis dataset. The C-Net model achieves 99.34\% which is 1.71\% higher than the highest average accuracy in \cite{sharma2020novel} among the preceding models. We compare the accuracies on all magnification factors in models that used a hybrid architecture. The highest accuracy is 97.2\% with an average of 96.48\% in \cite{pratiher2018grading}, the highest accuracy is 97.95\% with an average of 97.55\% achieved by \cite{alom2019breast}. Our model uses the same metrics and dataset with the values of 99.67\% and 99.34\% for the highest and average accuracy scores respectively. The performance shows that our architecture  with multiple networks as the feature extractor in Outer networks provide better features for the network to learn different variations in images. 

On the Osteosarcoma dataset, \autoref{tab:osteo results} shows that the least accuracy achieved by our model, in NCT vs NT classification categories, has a rate of 95.83\% which is 3.23\% higher than the best performance stated in \cite{arunachalam2019viable}. Moreover, comparing the best performance of our model with the 100\% accuracy on VT vs NT category, shows 7.4\% improvement.

The model achieves 100\% precision on two classification categories, NCT vs VT and VT vs NT, and the average of all three categories is 98.25\%. Similarly the model obtains 99.18\% of specificity, shown in \autoref{tab:osteo results}, which is 5.35\% and 3.08\% higher, respectively, compared to the results in \cite{fu2020deep}. Illustrated in \autoref{fig:confusionMat_Osteosarcoma}, the model predicts zero misclassification in detecting viable tumors versus non-tumors on the Osteosarcoma dataset, which clearly shows the ability and reliability of the proposed model for classification of biomedical images. 

\section{Conclusion} \label{conclusion}
In this work, we proposed a novel CNN, called C-Net, for binary classification of biomedical images. The architecture was tested on two different datasets, namely, BreakHis and Osteosarcoma, with regards to different image sizes, data volumes, and underlying diseases. To assure reliability, the proposed model has been evaluated using several metrics. The C-Net model achieved an average accuracy of 99.34\% and the average F1-score of 99.52\%, both of which outperformed all other models tested on the BreakHis dataset. In addition, the proposed model improved the accuracy and F1-score by 2.71\% and 4.76\%, respectively, compared to all the prior results on the Osteosarcoma dataset. The experimental results using a broad range of evaluation metrics showed that the proposed model is a promising architecture that has the potentials to be generalized on other biomedical image classification tasks.

\bibliographystyle{elsarticle-num-names}
\bibliography{References.bib}

\begin{thebibliography}{60}
\expandafter\ifx\csname natexlab\endcsname\relax\def\natexlab#1{#1}\fi
\providecommand{\url}[1]{\texttt{#1}}
\providecommand{\href}[2]{#2}
\providecommand{\path}[1]{#1}
\providecommand{\DOIprefix}{doi:}
\providecommand{\ArXivprefix}{arXiv:}
\providecommand{\URLprefix}{URL: }
\providecommand{\Pubmedprefix}{pmid:}
\providecommand{\doi}[1]{\href{http://dx.doi.org/#1}{\path{#1}}}
\providecommand{\Pubmed}[1]{\href{pmid:#1}{\path{#1}}}
\providecommand{\bibinfo}[2]{#2}
\ifx\xfnm\relax \def\xfnm[#1]{\unskip,\space#1}\fi
\bibitem[{Bray et~al.(2018)Bray, Ferlay, Soerjomataram, Siegel, Torre, and
  Jemal}]{cancerstat}
\bibinfo{author}{F.~Bray}, \bibinfo{author}{J.~Ferlay},
  \bibinfo{author}{I.~Soerjomataram}, \bibinfo{author}{R.~L. Siegel},
  \bibinfo{author}{L.~A. Torre}, \bibinfo{author}{A.~Jemal},
\newblock \bibinfo{title}{Global cancer statistics 2018: Globocan estimates of
  incidence and mortality worldwide for 36 cancers in 185 countries},
\newblock \bibinfo{journal}{CA: A Cancer Journal for Clinicians}
  \bibinfo{volume}{68} (\bibinfo{year}{2018}) \bibinfo{pages}{394--424}.
  \URLprefix
  \url{https://acsjournals.onlinelibrary.wiley.com/doi/abs/10.3322/caac.21492}.
  \DOIprefix\doi{10.3322/caac.21492}.
\bibitem[{Wild et~al.(2020)Wild, Weiderpass, and Stewart}]{wild2020world}
\bibinfo{author}{C.~Wild}, \bibinfo{author}{E.~Weiderpass},
  \bibinfo{author}{B.~Stewart},
\newblock \bibinfo{title}{World cancer report: cancer research for cancer
  prevention},
\newblock \bibinfo{journal}{Lyon, France: International Agency for Research on
  Cancer}  (\bibinfo{year}{2020}). \URLprefix
  \url{http://publications.iarc.fr/586}.
\bibitem[{Rubin and Reisner(2014)}]{rubin2014essentials}
\bibinfo{author}{E.~Rubin}, \bibinfo{author}{H.~M. Reisner},
  \bibinfo{title}{Essentials of Rubin's pathology Sixth Edition},
  \bibinfo{publisher}{Lippincott Williams \& Wilkins}, \bibinfo{year}{2014}.
\bibitem[{Rodwell et~al.(2015)Rodwell, Bender, Botham, Kennelly, and
  Weil}]{rodwell2015harpers}
\bibinfo{author}{V.~Rodwell}, \bibinfo{author}{D.~Bender},
  \bibinfo{author}{K.~M. Botham}, \bibinfo{author}{P.~J. Kennelly},
  \bibinfo{author}{P.~A. Weil}, \bibinfo{title}{Harpers illustrated
  biochemistry 30th edition}, \bibinfo{year}{2015}.
\bibitem[{Fass(2008)}]{fass2008imaging}
\bibinfo{author}{L.~Fass},
\newblock \bibinfo{title}{Imaging and cancer: a review},
\newblock \bibinfo{journal}{Molecular oncology} \bibinfo{volume}{2}
  (\bibinfo{year}{2008}) \bibinfo{pages}{115--152}.
\bibitem[{Prabhakar et~al.(2018)Prabhakar, Shende, and
  Augustine}]{prabhakar2018current}
\bibinfo{author}{B.~Prabhakar}, \bibinfo{author}{P.~Shende},
  \bibinfo{author}{S.~Augustine},
\newblock \bibinfo{title}{Current trends and emerging diagnostic techniques for
  lung cancer},
\newblock \bibinfo{journal}{Biomedicine \& Pharmacotherapy}
  \bibinfo{volume}{106} (\bibinfo{year}{2018}) \bibinfo{pages}{1586--1599}.
\bibitem[{Geller and Gorlick(2010)}]{geller2010osteosarcoma}
\bibinfo{author}{D.~S. Geller}, \bibinfo{author}{R.~Gorlick},
\newblock \bibinfo{title}{Osteosarcoma: a review of diagnosis, management, and
  treatment strategies},
\newblock \bibinfo{journal}{Clin Adv Hematol Oncol} \bibinfo{volume}{8}
  (\bibinfo{year}{2010}) \bibinfo{pages}{705--718}.
\bibitem[{Wang et~al.(2019)Wang, Yang, Rong, Zhan, and
  Xiao}]{wang2019pathology}
\bibinfo{author}{S.~Wang}, \bibinfo{author}{D.~M. Yang},
  \bibinfo{author}{R.~Rong}, \bibinfo{author}{X.~Zhan},
  \bibinfo{author}{G.~Xiao},
\newblock \bibinfo{title}{Pathology image analysis using segmentation deep
  learning algorithms},
\newblock \bibinfo{journal}{The American journal of pathology}
  \bibinfo{volume}{189} (\bibinfo{year}{2019}) \bibinfo{pages}{1686--1698}.
\bibitem[{Feng et~al.(2018)Feng, Zhang, and Mo}]{feng2018deep}
\bibinfo{author}{Y.~Feng}, \bibinfo{author}{L.~Zhang}, \bibinfo{author}{J.~Mo},
\newblock \bibinfo{title}{Deep manifold preserving autoencoder for classifying
  breast cancer histopathological images},
\newblock \bibinfo{journal}{IEEE/ACM Transactions on Computational Biology and
  Bioinformatics}  (\bibinfo{year}{2018}).
\bibitem[{Litjens et~al.(2016)Litjens, S{\'a}nchez, Timofeeva, Hermsen,
  Nagtegaal, Kovacs, Hulsbergen-Van De~Kaa, Bult, Van~Ginneken, and Van
  Der~Laak}]{litjens2016deep}
\bibinfo{author}{G.~Litjens}, \bibinfo{author}{C.~I. S{\'a}nchez},
  \bibinfo{author}{N.~Timofeeva}, \bibinfo{author}{M.~Hermsen},
  \bibinfo{author}{I.~Nagtegaal}, \bibinfo{author}{I.~Kovacs},
  \bibinfo{author}{C.~Hulsbergen-Van De~Kaa}, \bibinfo{author}{P.~Bult},
  \bibinfo{author}{B.~Van~Ginneken}, \bibinfo{author}{J.~Van Der~Laak},
\newblock \bibinfo{title}{Deep learning as a tool for increased accuracy and
  efficiency of histopathological diagnosis},
\newblock \bibinfo{journal}{Scientific reports} \bibinfo{volume}{6}
  (\bibinfo{year}{2016}) \bibinfo{pages}{26286}.
\bibitem[{Pantanowitz et~al.(2011)Pantanowitz, Valenstein, Evans, Kaplan,
  Pfeifer, Wilbur, Collins, and Colgan}]{pantanowitz2011review}
\bibinfo{author}{L.~Pantanowitz}, \bibinfo{author}{P.~N. Valenstein},
  \bibinfo{author}{A.~J. Evans}, \bibinfo{author}{K.~J. Kaplan},
  \bibinfo{author}{J.~D. Pfeifer}, \bibinfo{author}{D.~C. Wilbur},
  \bibinfo{author}{L.~C. Collins}, \bibinfo{author}{T.~J. Colgan},
\newblock \bibinfo{title}{Review of the current state of whole slide imaging in
  pathology},
\newblock \bibinfo{journal}{Journal of pathology informatics}
  \bibinfo{volume}{2} (\bibinfo{year}{2011}).
\bibitem[{Lisboa and Taktak(2006)}]{lisboa2006use}
\bibinfo{author}{P.~J. Lisboa}, \bibinfo{author}{A.~F. Taktak},
\newblock \bibinfo{title}{The use of artificial neural networks in decision
  support in cancer: a systematic review},
\newblock \bibinfo{journal}{Neural networks} \bibinfo{volume}{19}
  (\bibinfo{year}{2006}) \bibinfo{pages}{408--415}.
\bibitem[{Ibrahim et~al.(2020)Ibrahim, Gamble, Jaroensri, Abdelsamea, Mermel,
  Chen, and Rakha}]{ibrahim2020artificial}
\bibinfo{author}{A.~Ibrahim}, \bibinfo{author}{P.~Gamble},
  \bibinfo{author}{R.~Jaroensri}, \bibinfo{author}{M.~M. Abdelsamea},
  \bibinfo{author}{C.~H. Mermel}, \bibinfo{author}{P.-H.~C. Chen},
  \bibinfo{author}{E.~A. Rakha},
\newblock \bibinfo{title}{Artificial intelligence in digital breast pathology:
  Techniques and applications},
\newblock \bibinfo{journal}{The Breast} \bibinfo{volume}{49}
  (\bibinfo{year}{2020}) \bibinfo{pages}{267--273}.
\bibitem[{Schmidhuber(2015)}]{schmidhuber2015deep}
\bibinfo{author}{J.~Schmidhuber},
\newblock \bibinfo{title}{Deep learning in neural networks: An overview},
\newblock \bibinfo{journal}{Neural networks} \bibinfo{volume}{61}
  (\bibinfo{year}{2015}) \bibinfo{pages}{85--117}.
\bibitem[{Duggento et~al.(2020)Duggento, Conti, Mauriello, Guerrisi, and
  Toschi}]{duggento2020deep}
\bibinfo{author}{A.~Duggento}, \bibinfo{author}{A.~Conti},
  \bibinfo{author}{A.~Mauriello}, \bibinfo{author}{M.~Guerrisi},
  \bibinfo{author}{N.~Toschi},
\newblock \bibinfo{title}{Deep computational pathology in breast cancer},
\newblock in: \bibinfo{booktitle}{Seminars in Cancer Biology},
  \bibinfo{organization}{Elsevier}, \bibinfo{year}{2020}.
\bibitem[{Goodfellow et~al.(2016)Goodfellow, Bengio, Courville, and
  Bengio}]{goodfellow2016deep}
\bibinfo{author}{I.~Goodfellow}, \bibinfo{author}{Y.~Bengio},
  \bibinfo{author}{A.~Courville}, \bibinfo{author}{Y.~Bengio},
  \bibinfo{title}{Deep learning}, volume~\bibinfo{volume}{1},
  \bibinfo{publisher}{MIT press Cambridge}, \bibinfo{year}{2016}.
\bibitem[{Talo(2019)}]{Talo_2019}
\bibinfo{author}{M.~Talo},
\newblock \bibinfo{title}{Automated classification of histopathology images
  using transfer learning},
\newblock \bibinfo{journal}{Artificial Intelligence in Medicine}
  \bibinfo{volume}{101} (\bibinfo{year}{2019}) \bibinfo{pages}{101743}.
  \URLprefix \url{http://dx.doi.org/10.1016/j.artmed.2019.101743}.
  \DOIprefix\doi{10.1016/j.artmed.2019.101743}.
\bibitem[{Krizhevsky et~al.(2012)Krizhevsky, Sutskever, and
  Hinton}]{krizhevsky2012imagenet}
\bibinfo{author}{A.~Krizhevsky}, \bibinfo{author}{I.~Sutskever},
  \bibinfo{author}{G.~E. Hinton},
\newblock \bibinfo{title}{Imagenet classification with deep convolutional
  neural networks},
\newblock in: \bibinfo{booktitle}{Advances in neural information processing
  systems}, \bibinfo{year}{2012}, pp. \bibinfo{pages}{1097--1105}.
\bibitem[{Lin et~al.(2013)Lin, Chen, and Yan}]{lin2013network}
\bibinfo{author}{M.~Lin}, \bibinfo{author}{Q.~Chen}, \bibinfo{author}{S.~Yan},
\newblock \bibinfo{title}{Network in network},
\newblock \bibinfo{journal}{arXiv preprint arXiv:1312.4400}
  (\bibinfo{year}{2013}).
\bibitem[{Szegedy et~al.(2015)Szegedy, Liu, Jia, Sermanet, Reed, Anguelov,
  Erhan, Vanhoucke, and Rabinovich}]{szegedy2015going}
\bibinfo{author}{C.~Szegedy}, \bibinfo{author}{W.~Liu},
  \bibinfo{author}{Y.~Jia}, \bibinfo{author}{P.~Sermanet},
  \bibinfo{author}{S.~Reed}, \bibinfo{author}{D.~Anguelov},
  \bibinfo{author}{D.~Erhan}, \bibinfo{author}{V.~Vanhoucke},
  \bibinfo{author}{A.~Rabinovich},
\newblock \bibinfo{title}{Going deeper with convolutions},
\newblock in: \bibinfo{booktitle}{Proceedings of the IEEE conference on
  computer vision and pattern recognition}, \bibinfo{year}{2015}, pp.
  \bibinfo{pages}{1--9}.
\bibitem[{Litjens et~al.(2017)Litjens, Kooi, Bejnordi, Setio, Ciompi,
  Ghafoorian, Van Der~Laak, Van~Ginneken, and S{\'a}nchez}]{litjens2017survey}
\bibinfo{author}{G.~Litjens}, \bibinfo{author}{T.~Kooi}, \bibinfo{author}{B.~E.
  Bejnordi}, \bibinfo{author}{A.~A.~A. Setio}, \bibinfo{author}{F.~Ciompi},
  \bibinfo{author}{M.~Ghafoorian}, \bibinfo{author}{J.~A. Van Der~Laak},
  \bibinfo{author}{B.~Van~Ginneken}, \bibinfo{author}{C.~I. S{\'a}nchez},
\newblock \bibinfo{title}{A survey on deep learning in medical image analysis},
\newblock \bibinfo{journal}{Medical image analysis} \bibinfo{volume}{42}
  (\bibinfo{year}{2017}) \bibinfo{pages}{60--88}.
\bibitem[{Esteva et~al.(2017)Esteva, Kuprel, Novoa, Ko, Swetter, Blau, and
  Thrun}]{esteva2017dermatologist}
\bibinfo{author}{A.~Esteva}, \bibinfo{author}{B.~Kuprel},
  \bibinfo{author}{R.~A. Novoa}, \bibinfo{author}{J.~Ko},
  \bibinfo{author}{S.~M. Swetter}, \bibinfo{author}{H.~M. Blau},
  \bibinfo{author}{S.~Thrun},
\newblock \bibinfo{title}{Dermatologist-level classification of skin cancer
  with deep neural networks},
\newblock \bibinfo{journal}{nature} \bibinfo{volume}{542}
  (\bibinfo{year}{2017}) \bibinfo{pages}{115--118}.
\bibitem[{Zhang et~al.(2019)Zhang, Sun, Dang, Li, Guo, Chang, Yu, Huang, Wu,
  Liang et~al.}]{zhang2019toward}
\bibinfo{author}{C.~Zhang}, \bibinfo{author}{X.~Sun},
  \bibinfo{author}{K.~Dang}, \bibinfo{author}{K.~Li}, \bibinfo{author}{X.-w.
  Guo}, \bibinfo{author}{J.~Chang}, \bibinfo{author}{Z.-q. Yu},
  \bibinfo{author}{F.-y. Huang}, \bibinfo{author}{Y.-s. Wu},
  \bibinfo{author}{Z.~Liang}, et~al.,
\newblock \bibinfo{title}{Toward an expert level of lung cancer detection and
  classification using a deep convolutional neural network},
\newblock \bibinfo{journal}{The oncologist} \bibinfo{volume}{24}
  (\bibinfo{year}{2019}) \bibinfo{pages}{1159}.
\bibitem[{Cruz-Roa et~al.(2017)Cruz-Roa, Gilmore, Basavanhally, Feldman,
  Ganesan, Shih, Tomaszewski, Gonz{\'a}lez, and Madabhushi}]{cruz2017accurate}
\bibinfo{author}{A.~Cruz-Roa}, \bibinfo{author}{H.~Gilmore},
  \bibinfo{author}{A.~Basavanhally}, \bibinfo{author}{M.~Feldman},
  \bibinfo{author}{S.~Ganesan}, \bibinfo{author}{N.~N. Shih},
  \bibinfo{author}{J.~Tomaszewski}, \bibinfo{author}{F.~A. Gonz{\'a}lez},
  \bibinfo{author}{A.~Madabhushi},
\newblock \bibinfo{title}{Accurate and reproducible invasive breast cancer
  detection in whole-slide images: A deep learning approach for quantifying
  tumor extent},
\newblock \bibinfo{journal}{Scientific reports} \bibinfo{volume}{7}
  (\bibinfo{year}{2017}) \bibinfo{pages}{46450}.
\bibitem[{{Chang} et~al.(2017){Chang}, {Yu}, {Han}, {Chang}, and
  {Park}}]{chang}
\bibinfo{author}{J.~{Chang}}, \bibinfo{author}{J.~{Yu}},
  \bibinfo{author}{T.~{Han}}, \bibinfo{author}{H.~{Chang}},
  \bibinfo{author}{E.~{Park}},
\newblock \bibinfo{title}{A method for classifying medical images using
  transfer learning: A pilot study on histopathology of breast cancer},
\newblock in: \bibinfo{booktitle}{2017 IEEE 19th International Conference on
  e-Health Networking, Applications and Services (Healthcom)},
  \bibinfo{year}{2017}, pp. \bibinfo{pages}{1--4}.
\bibitem[{Wahab et~al.(2017)Wahab, Khan, and Lee}]{wahab2017two}
\bibinfo{author}{N.~Wahab}, \bibinfo{author}{A.~Khan}, \bibinfo{author}{Y.~S.
  Lee},
\newblock \bibinfo{title}{Two-phase deep convolutional neural network for
  reducing class skewness in histopathological images based breast cancer
  detection},
\newblock \bibinfo{journal}{Computers in biology and medicine}
  \bibinfo{volume}{85} (\bibinfo{year}{2017}) \bibinfo{pages}{86--97}.
\bibitem[{Roy et~al.(2019)Roy, Banik, Bhattacharjee, and
  Nasipuri}]{roy2019patch}
\bibinfo{author}{K.~Roy}, \bibinfo{author}{D.~Banik},
  \bibinfo{author}{D.~Bhattacharjee}, \bibinfo{author}{M.~Nasipuri},
\newblock \bibinfo{title}{Patch-based system for classification of breast
  histology images using deep learning},
\newblock \bibinfo{journal}{Computerized Medical Imaging and Graphics}
  \bibinfo{volume}{71} (\bibinfo{year}{2019}) \bibinfo{pages}{90--103}.
\bibitem[{Fond{\'o}n et~al.(2018)Fond{\'o}n, Sarmiento, Garc{\'\i}a, Silvestre,
  Eloy, Pol{\'o}nia, and Aguiar}]{fondon2018automatic}
\bibinfo{author}{I.~Fond{\'o}n}, \bibinfo{author}{A.~Sarmiento},
  \bibinfo{author}{A.~I. Garc{\'\i}a}, \bibinfo{author}{M.~Silvestre},
  \bibinfo{author}{C.~Eloy}, \bibinfo{author}{A.~Pol{\'o}nia},
  \bibinfo{author}{P.~Aguiar},
\newblock \bibinfo{title}{Automatic classification of tissue malignancy for
  breast carcinoma diagnosis},
\newblock \bibinfo{journal}{Computers in biology and medicine}
  \bibinfo{volume}{96} (\bibinfo{year}{2018}) \bibinfo{pages}{41--51}.
\bibitem[{Han et~al.(2017)Han, Wei, Zheng, Yin, Li, and Li}]{han2017breast}
\bibinfo{author}{Z.~Han}, \bibinfo{author}{B.~Wei}, \bibinfo{author}{Y.~Zheng},
  \bibinfo{author}{Y.~Yin}, \bibinfo{author}{K.~Li}, \bibinfo{author}{S.~Li},
\newblock \bibinfo{title}{Breast cancer multi-classification from
  histopathological images with structured deep learning model},
\newblock \bibinfo{journal}{Scientific reports} \bibinfo{volume}{7}
  (\bibinfo{year}{2017}) \bibinfo{pages}{1--10}.
\bibitem[{Xu et~al.(2017)Xu, Chen, Pei, Chang, and Du}]{xu2017effect}
\bibinfo{author}{Y.~X. Xu}, \bibinfo{author}{L.~Chen},
  \bibinfo{author}{F.~Pei}, \bibinfo{author}{K.~K. Chang},
  \bibinfo{author}{Y.~Du},
\newblock \bibinfo{title}{Effect of the modulation ratio on the interface
  structure of tialn/tin and tialn/zrn multilayers: first-principles and
  experimental investigations},
\newblock \bibinfo{journal}{Acta Materialia} \bibinfo{volume}{130}
  (\bibinfo{year}{2017}) \bibinfo{pages}{281--288}.
\bibitem[{Talo(2019)}]{talo2019automated}
\bibinfo{author}{M.~Talo},
\newblock \bibinfo{title}{Automated classification of histopathology images
  using transfer learning},
\newblock \bibinfo{journal}{Artificial Intelligence in Medicine}
  \bibinfo{volume}{101} (\bibinfo{year}{2019}) \bibinfo{pages}{101743}.
\bibitem[{de~Matos et~al.(2019)de~Matos, Britto, Oliveira, and
  Koerich}]{de2019double}
\bibinfo{author}{J.~de~Matos}, \bibinfo{author}{A.~d.~S. Britto},
  \bibinfo{author}{L.~E. Oliveira}, \bibinfo{author}{A.~L. Koerich},
\newblock \bibinfo{title}{Double transfer learning for breast cancer
  histopathologic image classification},
\newblock in: \bibinfo{booktitle}{2019 International Joint Conference on Neural
  Networks (IJCNN)}, \bibinfo{organization}{IEEE}, \bibinfo{year}{2019}, pp.
  \bibinfo{pages}{1--8}.
\bibitem[{Celik et~al.(2020)Celik, Talo, Yildirim, Karabatak, and
  Acharya}]{celik2020automated}
\bibinfo{author}{Y.~Celik}, \bibinfo{author}{M.~Talo},
  \bibinfo{author}{O.~Yildirim}, \bibinfo{author}{M.~Karabatak},
  \bibinfo{author}{U.~R. Acharya},
\newblock \bibinfo{title}{Automated invasive ductal carcinoma detection based
  using deep transfer learning with whole-slide images},
\newblock \bibinfo{journal}{Pattern Recognition Letters}
  (\bibinfo{year}{2020}).
\bibitem[{Pratiher et~al.(2018)Pratiher, Chattoraj, Agarwal, and
  Bhattacharya}]{pratiher2018grading}
\bibinfo{author}{S.~Pratiher}, \bibinfo{author}{S.~Chattoraj},
  \bibinfo{author}{S.~Agarwal}, \bibinfo{author}{S.~Bhattacharya},
\newblock \bibinfo{title}{Grading tumor malignancy via deep bidirectional lstm
  on graph manifold encoded histopathological image},
\newblock in: \bibinfo{booktitle}{2018 IEEE International Conference on Data
  Mining Workshops (ICDMW)}, \bibinfo{organization}{IEEE},
  \bibinfo{year}{2018}, pp. \bibinfo{pages}{674--681}.
\bibitem[{Ara{\'u}jo et~al.(2017)Ara{\'u}jo, Aresta, Castro, Rouco, Aguiar,
  Eloy, Pol{\'o}nia, and Campilho}]{araujo2017classification}
\bibinfo{author}{T.~Ara{\'u}jo}, \bibinfo{author}{G.~Aresta},
  \bibinfo{author}{E.~Castro}, \bibinfo{author}{J.~Rouco},
  \bibinfo{author}{P.~Aguiar}, \bibinfo{author}{C.~Eloy},
  \bibinfo{author}{A.~Pol{\'o}nia}, \bibinfo{author}{A.~Campilho},
\newblock \bibinfo{title}{Classification of breast cancer histology images
  using convolutional neural networks},
\newblock \bibinfo{journal}{PloS one} \bibinfo{volume}{12}
  (\bibinfo{year}{2017}) \bibinfo{pages}{e0177544}.
\bibitem[{Spanhol et~al.(2017)Spanhol, Oliveira, Cavalin, Petitjean, and
  Heutte}]{spanhol2017deep}
\bibinfo{author}{F.~A. Spanhol}, \bibinfo{author}{L.~S. Oliveira},
  \bibinfo{author}{P.~R. Cavalin}, \bibinfo{author}{C.~Petitjean},
  \bibinfo{author}{L.~Heutte},
\newblock \bibinfo{title}{Deep features for breast cancer histopathological
  image classification},
\newblock in: \bibinfo{booktitle}{2017 IEEE International Conference on
  Systems, Man, and Cybernetics (SMC)}, \bibinfo{organization}{IEEE},
  \bibinfo{year}{2017}, pp. \bibinfo{pages}{1868--1873}.
\bibitem[{Deniz et~al.(2018)Deniz, {\c{S}}eng{\"u}r, Kadiro{\u{g}}lu, Guo,
  Bajaj, and Budak}]{deniz2018transfer}
\bibinfo{author}{E.~Deniz}, \bibinfo{author}{A.~{\c{S}}eng{\"u}r},
  \bibinfo{author}{Z.~Kadiro{\u{g}}lu}, \bibinfo{author}{Y.~Guo},
  \bibinfo{author}{V.~Bajaj}, \bibinfo{author}{{\"U}.~Budak},
\newblock \bibinfo{title}{Transfer learning based histopathologic image
  classification for breast cancer detection},
\newblock \bibinfo{journal}{Health information science and systems}
  \bibinfo{volume}{6} (\bibinfo{year}{2018}) \bibinfo{pages}{18}.
\bibitem[{Awan et~al.(2018)Awan, Koohbanani, Shaban, Lisowska, and
  Rajpoot}]{awan2018context}
\bibinfo{author}{R.~Awan}, \bibinfo{author}{N.~A. Koohbanani},
  \bibinfo{author}{M.~Shaban}, \bibinfo{author}{A.~Lisowska},
  \bibinfo{author}{N.~Rajpoot},
\newblock \bibinfo{title}{Context-aware learning using transferable features
  for classification of breast cancer histology images},
\newblock in: \bibinfo{booktitle}{International Conference Image Analysis and
  Recognition}, \bibinfo{organization}{Springer}, \bibinfo{year}{2018}, pp.
  \bibinfo{pages}{788--795}.
\bibitem[{Wahab et~al.(2019)Wahab, Khan, and Lee}]{wahab2019transfer}
\bibinfo{author}{N.~Wahab}, \bibinfo{author}{A.~Khan}, \bibinfo{author}{Y.~S.
  Lee},
\newblock \bibinfo{title}{Transfer learning based deep cnn for segmentation and
  detection of mitoses in breast cancer histopathological images},
\newblock \bibinfo{journal}{Microscopy} \bibinfo{volume}{68}
  (\bibinfo{year}{2019}) \bibinfo{pages}{216--233}.
\bibitem[{Khan et~al.(2019)Khan, Islam, Jan, Din, and
  Rodrigues}]{khan2019novel}
\bibinfo{author}{S.~Khan}, \bibinfo{author}{N.~Islam},
  \bibinfo{author}{Z.~Jan}, \bibinfo{author}{I.~U. Din},
  \bibinfo{author}{J.~J.~C. Rodrigues},
\newblock \bibinfo{title}{A novel deep learning based framework for the
  detection and classification of breast cancer using transfer learning},
\newblock \bibinfo{journal}{Pattern Recognition Letters} \bibinfo{volume}{125}
  (\bibinfo{year}{2019}) \bibinfo{pages}{1--6}.
\bibitem[{Gupta and Bhavsar(2019)}]{gupta2019partially}
\bibinfo{author}{V.~Gupta}, \bibinfo{author}{A.~Bhavsar},
\newblock \bibinfo{title}{Partially-independent framework for breast cancer
  histopathological image classification},
\newblock in: \bibinfo{booktitle}{Proceedings of the IEEE Conference on
  Computer Vision and Pattern Recognition Workshops}, \bibinfo{year}{2019}, pp.
  \bibinfo{pages}{0--0}.
\bibitem[{Arunachalam et~al.(2019)Arunachalam, Mishra, Daescu, Cederberg,
  Rakheja, Sengupta, Leonard, Hallac, and Leavey}]{arunachalam2019viable}
\bibinfo{author}{H.~B. Arunachalam}, \bibinfo{author}{R.~Mishra},
  \bibinfo{author}{O.~Daescu}, \bibinfo{author}{K.~Cederberg},
  \bibinfo{author}{D.~Rakheja}, \bibinfo{author}{A.~Sengupta},
  \bibinfo{author}{D.~Leonard}, \bibinfo{author}{R.~Hallac},
  \bibinfo{author}{P.~Leavey},
\newblock \bibinfo{title}{Viable and necrotic tumor assessment from whole slide
  images of osteosarcoma using machine-learning and deep-learning models},
\newblock \bibinfo{journal}{PloS one} \bibinfo{volume}{14}
  (\bibinfo{year}{2019}) \bibinfo{pages}{e0210706}.
\bibitem[{Fu et~al.(2020)Fu, Xue, Ji, Cui, and Dong}]{fu2020deep}
\bibinfo{author}{Y.~Fu}, \bibinfo{author}{P.~Xue}, \bibinfo{author}{H.~Ji},
  \bibinfo{author}{W.~Cui}, \bibinfo{author}{E.~Dong},
\newblock \bibinfo{title}{Deep model with siamese network for viable and
  necrotic tumor regions assessment in osteosarcoma},
\newblock \bibinfo{journal}{Medical Physics}  (\bibinfo{year}{2020}).
\bibitem[{Anisuzzaman et~al.(2021)Anisuzzaman, Barzekar, Tong, Luo, and
  Yu}]{ANISUZZAMAN2021102931}
\bibinfo{author}{D.~Anisuzzaman}, \bibinfo{author}{H.~Barzekar},
  \bibinfo{author}{L.~Tong}, \bibinfo{author}{J.~Luo}, \bibinfo{author}{Z.~Yu},
\newblock \bibinfo{title}{A deep learning study on osteosarcoma detection from
  histological images},
\newblock \bibinfo{journal}{Biomedical Signal Processing and Control}
  \bibinfo{volume}{69} (\bibinfo{year}{2021}) \bibinfo{pages}{102931}.
  \URLprefix
  \url{https://www.sciencedirect.com/science/article/pii/S1746809421005280}.
  \DOIprefix\doi{https://doi.org/10.1016/j.bspc.2021.102931}.
\bibitem[{Spanhol et~al.(2015)Spanhol, Oliveira, Petitjean, and
  Heutte}]{spanhol2015dataset}
\bibinfo{author}{F.~A. Spanhol}, \bibinfo{author}{L.~S. Oliveira},
  \bibinfo{author}{C.~Petitjean}, \bibinfo{author}{L.~Heutte},
\newblock \bibinfo{title}{A dataset for breast cancer histopathological image
  classification},
\newblock \bibinfo{journal}{IEEE Transactions on Biomedical Engineering}
  \bibinfo{volume}{63} (\bibinfo{year}{2015}) \bibinfo{pages}{1455--1462}.
\bibitem[{The Cancer Imaging~Archive(2019)}]{osteodataset}
\bibinfo{author}{T.~The Cancer Imaging~Archive}, \bibinfo{title}{Osteosarcoma
  data from ut southwestern ut dallas for viable and necrotic tumor
  assessment}, \bibinfo{year}{2019}. \URLprefix
  \url{https://doi.org/10.7937/tcia.2019.bvhjhdas.}
\bibitem[{Shorten and Khoshgoftaar(2019)}]{shorten2019survey}
\bibinfo{author}{C.~Shorten}, \bibinfo{author}{T.~M. Khoshgoftaar},
\newblock \bibinfo{title}{A survey on image data augmentation for deep
  learning},
\newblock \bibinfo{journal}{Journal of Big Data} \bibinfo{volume}{6}
  (\bibinfo{year}{2019}) \bibinfo{pages}{60}.
\bibitem[{Simonyan and Zisserman(2014)}]{simonyan2014very}
\bibinfo{author}{K.~Simonyan}, \bibinfo{author}{A.~Zisserman},
\newblock \bibinfo{title}{Very deep convolutional networks for large-scale
  image recognition},
\newblock \bibinfo{journal}{arXiv preprint arXiv:1409.1556}
  (\bibinfo{year}{2014}).
\bibitem[{Abadi et~al.(2016)Abadi, Barham, Chen, Chen, Davis, Dean, Devin,
  Ghemawat, Irving, Isard et~al.}]{abadi2016tensorflow}
\bibinfo{author}{M.~Abadi}, \bibinfo{author}{P.~Barham},
  \bibinfo{author}{J.~Chen}, \bibinfo{author}{Z.~Chen},
  \bibinfo{author}{A.~Davis}, \bibinfo{author}{J.~Dean},
  \bibinfo{author}{M.~Devin}, \bibinfo{author}{S.~Ghemawat},
  \bibinfo{author}{G.~Irving}, \bibinfo{author}{M.~Isard}, et~al.,
\newblock \bibinfo{title}{Tensorflow: A system for large-scale machine
  learning},
\newblock in: \bibinfo{booktitle}{12th $\{$USENIX$\}$ symposium on operating
  systems design and implementation ($\{$OSDI$\}$ 16)}, \bibinfo{year}{2016},
  pp. \bibinfo{pages}{265--283}.
\bibitem[{Chollet et~al.(2015)}]{chollet2015keras}
\bibinfo{author}{F.~Chollet}, et~al., \bibinfo{title}{Keras},
  \bibinfo{howpublished}{\url{https://github.com/fchollet/keras}},
  \bibinfo{year}{2015}.
\bibitem[{Chicco and Jurman(2020)}]{chicco2020advantages}
\bibinfo{author}{D.~Chicco}, \bibinfo{author}{G.~Jurman},
\newblock \bibinfo{title}{The advantages of the matthews correlation
  coefficient (mcc) over f1 score and accuracy in binary classification
  evaluation},
\newblock \bibinfo{journal}{BMC genomics} \bibinfo{volume}{21}
  (\bibinfo{year}{2020}) \bibinfo{pages}{6}.
\bibitem[{Alom et~al.(2019)Alom, Yakopcic, Nasrin, Taha, and
  Asari}]{alom2019breast}
\bibinfo{author}{M.~Z. Alom}, \bibinfo{author}{C.~Yakopcic},
  \bibinfo{author}{M.~S. Nasrin}, \bibinfo{author}{T.~M. Taha},
  \bibinfo{author}{V.~K. Asari},
\newblock \bibinfo{title}{Breast cancer classification from histopathological
  images with inception recurrent residual convolutional neural network},
\newblock \bibinfo{journal}{Journal of digital imaging} \bibinfo{volume}{32}
  (\bibinfo{year}{2019}) \bibinfo{pages}{605--617}.
\bibitem[{To{\u{g}}a{\c{c}}ar et~al.(2020)To{\u{g}}a{\c{c}}ar, {\"O}zkurt,
  Ergen, and C{\"o}mert}]{tougaccar2020breastnet}
\bibinfo{author}{M.~To{\u{g}}a{\c{c}}ar}, \bibinfo{author}{K.~B. {\"O}zkurt},
  \bibinfo{author}{B.~Ergen}, \bibinfo{author}{Z.~C{\"o}mert},
\newblock \bibinfo{title}{Breastnet: A novel convolutional neural network model
  through histopathological images for the diagnosis of breast cancer},
\newblock \bibinfo{journal}{Physica A: Statistical Mechanics and its
  Applications} \bibinfo{volume}{545} (\bibinfo{year}{2020})
  \bibinfo{pages}{123592}.
\bibitem[{Sharma et~al.(2020)Sharma, Verma, Mishra, and
  Bhattacharya}]{sharma2020novel}
\bibinfo{author}{M.~Sharma}, \bibinfo{author}{R.~Verma},
  \bibinfo{author}{A.~Mishra}, \bibinfo{author}{M.~Bhattacharya},
\newblock \bibinfo{title}{A novel approach to classify breast cancer tumors
  using deep learning approach and resulting most accurate magnification
  factor},
\newblock in: \bibinfo{booktitle}{High Performance Vision Intelligence},
  \bibinfo{publisher}{Springer}, \bibinfo{year}{2020}, pp.
  \bibinfo{pages}{185--201}.
\bibitem[{{Sun} et~al.(2020){Sun}, {Xu}, {Liu}, {Xiong}, {Zhao}, and
  {Ding}}]{sun2020deep}
\bibinfo{author}{C.~{Sun}}, \bibinfo{author}{A.~{Xu}},
  \bibinfo{author}{D.~{Liu}}, \bibinfo{author}{Z.~{Xiong}},
  \bibinfo{author}{F.~{Zhao}}, \bibinfo{author}{W.~{Ding}},
\newblock \bibinfo{title}{Deep learning-based classification of liver cancer
  histopathology images using only global labels},
\newblock \bibinfo{journal}{IEEE Journal of Biomedical and Health Informatics}
  \bibinfo{volume}{24} (\bibinfo{year}{2020}) \bibinfo{pages}{1643--1651}.
\bibitem[{Azer(2019)}]{azer2019challenges}
\bibinfo{author}{S.~A. Azer},
\newblock \bibinfo{title}{Challenges facing the detection of colonic polyps:
  What can deep learning do?},
\newblock \bibinfo{journal}{Medicina} \bibinfo{volume}{55}
  (\bibinfo{year}{2019}) \bibinfo{pages}{473}.
\bibitem[{Zheng et~al.(2020)Zheng, Yao, Huang, Yu, Wang, Liu, Mao, Li, Xiao,
  Wang et~al.}]{zheng2020deep}
\bibinfo{author}{X.~Zheng}, \bibinfo{author}{Z.~Yao},
  \bibinfo{author}{Y.~Huang}, \bibinfo{author}{Y.~Yu},
  \bibinfo{author}{Y.~Wang}, \bibinfo{author}{Y.~Liu},
  \bibinfo{author}{R.~Mao}, \bibinfo{author}{F.~Li}, \bibinfo{author}{Y.~Xiao},
  \bibinfo{author}{Y.~Wang}, et~al.,
\newblock \bibinfo{title}{Deep learning radiomics can predict axillary lymph
  node status in early-stage breast cancer},
\newblock \bibinfo{journal}{Nature communications} \bibinfo{volume}{11}
  (\bibinfo{year}{2020}) \bibinfo{pages}{1--9}.
\bibitem[{Chougrad et~al.(2018)Chougrad, Zouaki, and
  Alheyane}]{chougrad2018deep}
\bibinfo{author}{H.~Chougrad}, \bibinfo{author}{H.~Zouaki},
  \bibinfo{author}{O.~Alheyane},
\newblock \bibinfo{title}{Deep convolutional neural networks for breast cancer
  screening},
\newblock \bibinfo{journal}{Computer methods and programs in biomedicine}
  \bibinfo{volume}{157} (\bibinfo{year}{2018}) \bibinfo{pages}{19--30}.
\bibitem[{Lalkhen and McCluskey(2008)}]{lalkhen2008clinical}
\bibinfo{author}{A.~G. Lalkhen}, \bibinfo{author}{A.~McCluskey},
\newblock \bibinfo{title}{Clinical tests: sensitivity and specificity},
\newblock \bibinfo{journal}{Continuing Education in Anaesthesia Critical Care
  \& Pain} \bibinfo{volume}{8} (\bibinfo{year}{2008})
  \bibinfo{pages}{221--223}.
\bibitem[{Akobeng(2007)}]{akobeng2007understanding}
\bibinfo{author}{A.~K. Akobeng},
\newblock \bibinfo{title}{Understanding diagnostic tests 1: sensitivity,
  specificity and predictive values},
\newblock \bibinfo{journal}{Acta paediatrica} \bibinfo{volume}{96}
  (\bibinfo{year}{2007}) \bibinfo{pages}{338--341}.

\end{thebibliography}







\end{document}